\documentclass{PoS}

\newcommand{\vek}[1] {\boldsymbol{#1}}

\newcommand{\y}{\vek{y}}

\usepackage{amsmath,graphicx,verbatim}
\usepackage{epstopdf}
\usepackage[utf8x]{inputenc}
\usepackage{amsmath,graphicx}
\usepackage{simplewick}
\usepackage{datetime}
\usepackage{here}
\usepackage{setspace}

\usepackage{paralist}

\title{Polarization in double parton scattering}

\ShortTitle{Polarization in double parton scattering}

\author{\speaker{Tomas KASEMETS}%
\\
       Nikhef Theory Group and VU University Amsterdam, \\De Boelelaan 1081, NL-1081 HV Amsterdam, the Netherlands\\
       E-mail: \email{kasemets@nikhef.nl}}

\abstract{We briefly discuss the status of double parton scattering (DPS) on both theoretical and experimental sides and summarize the state of the naive DPS model, where the scatterers are taken completely independent. We then elaborate on some of the effects which are neglected in such an approach, and often ignored in phenomenological studies, with focus on polarization - which arises from the correlation between the spin of two partons inside a proton. Polarization is of particular interest, thanks to its direct and calculable connections to the distribution of particles in the final state. Although the physics described is different, there are strong similarities between polarization in DPS and in single partons scattering with measured transverse momentum, and several of the techniques and results from studies of transverse momentum dependent parton distributions can be translated into the setting of double parton scattering.}

\FullConference{XXIII International Workshop on Deep-Inelastic Scattering\\
		 27 April - May 1 2015\\
		 Dallas, Texas}

\begin{document}

\section{Brief introduction to double partons scattering}
Double parton scattering (DPS) describes collisions where two partons from each of two colliding protons interact in two separate hard collisions. There is a strong dependence on the amount of DPS in proton collisions on the density of partons inside the protons. As the density increases towards small momentum fractions, and as higher energy of the collider implies that smaller momentum fractions are probed, the relevance of DPS increases with collider energy. Hence, DPS has played a larger role at the LHC than at previous colliders, and is further enhanced with the restart of the LHC at higher center of mass energy.

This has inspired an increase of the activity in the field over recent years, and a large increase in the number of written papers on double parton scattering. Naturally, this has also lead to new insights, advanced the field towards more solid ground both theoretically and experimentally, but also led to new questions and problems that must be answered and solved.

Double partons scattering contributes to both signal and backgrounds of many final states under study at the LHC. So, if DPS is enhanced at the LHC, how come we do not take it into account in every cross section calculation? The reason is, that for inclusive enough observables, double parton scattering cancels in the sum over final states. Considering for example the production of a Higgs-boson, in the calculation $pp \rightarrow H + X$, double parton scattering affects $X$. As long as one remains inclusive enough and do not ask specific questions about $X$, the effects of double parton scattering will cancel thanks to unitarity. 

If we do start asking questions about $X$, for example asking for another Higgs-boson or dijet, we do get a contribution to the cross section where the second Higgs boson or dijet is produced in a second hard scattering. Another example is if one starts asking questions about observables where a second hard scattering can have an impact on the distribution, such as the transverse energy of the produced system \cite{Gaunt:2014ska}. Generically, we need to worry about DPS whenever we have 
\begin{compactitem}
\item final states with several particles (typically $\geq 4$)
\item high energy hadron collisions, where small momentum fractions are probed
\item and/or if single parton scattering is suppressed\
\end{compactitem}
These conditions are often fulfilled for studies at the LHC. Some examples are; double vector boson production - where same sign double $W$ is the cleanest DPS signal, double open charm ($D_0D_0$), double $J/\Psi$, $W+b$, 4 jets, photon + 3 jets etc.

The double parton scattering cross section can schematically be factorized as a product of two partonic cross sections convoluted with two double parton distributions (DPDs), 
\begin{align}\label{eq:x-sec}
\frac{d\sigma_{DPS}}{dx_id\bar{x}_i} \sim \hat{\sigma}_1(x_1\bar{x}_1) \hat{\sigma}_2(x_2\bar{x}_2) \int d^2\y F_{ab}(x_1,x_2,\y) \bar{F}_{cd}(\bar{x}_1,\bar{x}_2,\y).
\end{align}
The DPDs do not only depend on the flavor of the partons and their longitudinal momentum fractions $x_i$, but in addition on their transverse separation $\y$.  

There are quite a few things which \eqref{eq:x-sec} does not make explicit, and we will return to this point below. Before that, we want to briefly discuss the most simple approach to double parton scattering. If we start from \eqref{eq:x-sec}, forget about possible complications arising from what is kept implicit in this schematic formula, and then assume that the $\y$ dependence of the double parton distributions can be separated from the $x_i$ dependence, that the $\y$ dependence is universal (the same for all partons) and that the $x_i$ dependence is a product of two single parton distributions (PDFs). Then, the DPS cross section take an extremely simple form $\sigma_{DPS} \sim \sigma_1 \sigma_2/ \sigma_{eff}$ where $\sigma_{eff}$ parametrizes the integration over the transverse separation between the two hard interactions $\y$. This naive approximation for DPS, although in some cases unrealistic, has its clear virtues. If we know the single parton scattering cross section for process $1$ and $2$, then all that is standing between us and the determination of the DPS cross section is the knowledge of one number, $\sigma_{eff}$. This formula easily and directly makes the connection between phenomenology and experiment, and gives useful order of magnitude estimates for the DPS cross section.

\begin{figure}[t]
\begin{center}
\includegraphics[width=0.85\textwidth]{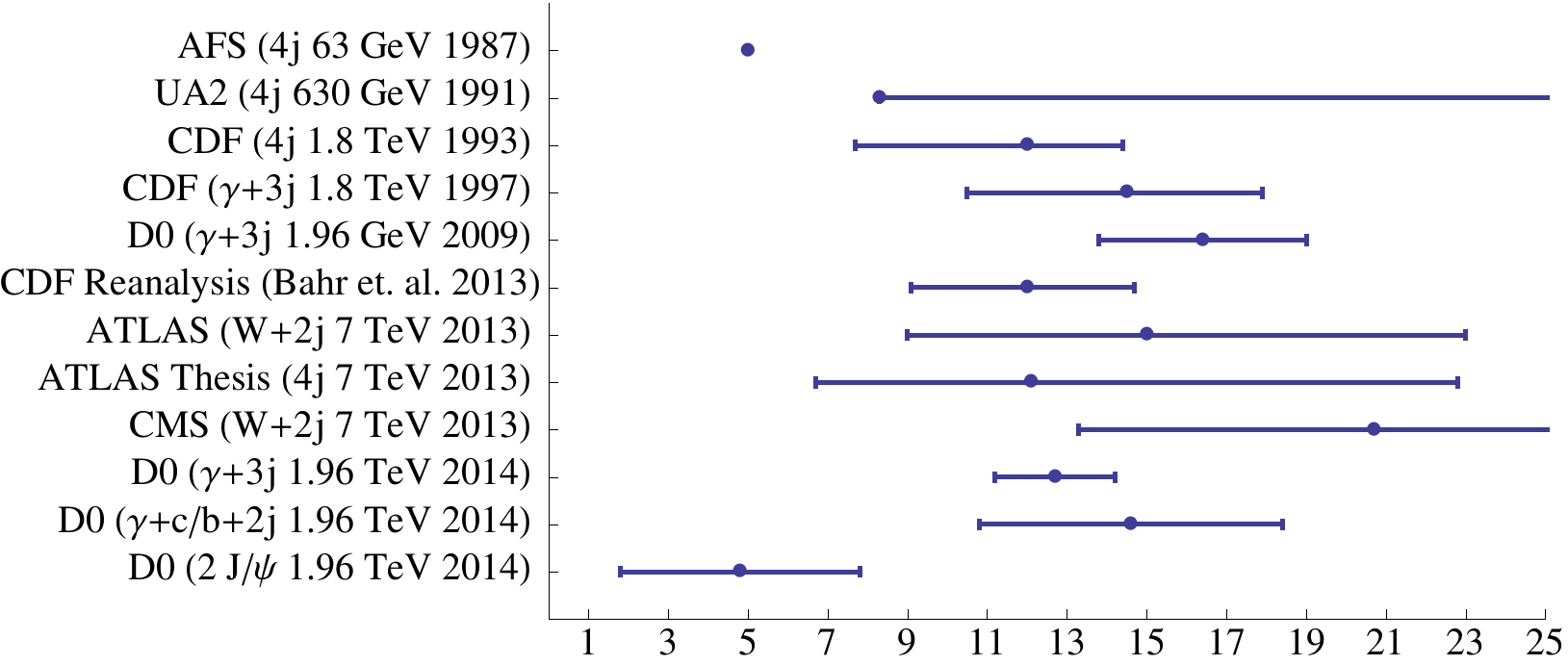}
\end{center}
\caption{\it
 Experimental extractions $\sigma_{\text{eff}}$ \cite{Akesson:1986iv,Alitti:1991rd,Abe:1993rv,Abe:1997xk,Abazov:2009gc,Aad:2013bjm,Chatrchyan:2013xxa,Bahr:2013gkj,Sadeh:2013wka,Abazov:2014fha,Abazov:2014qba}. Note that the AFS extraction does not have errors, and the UA2 extraction is a lower limit.
}
\label{fig:sigma_eff}
\end{figure}

Another point of view to look at this simple formula is that $\sigma_{eff}$ is parametrizing our ignorance, and by experimentally extracting $\sigma_{eff}$ in different processes, at different energies, rapidities etc, we can get indications on how well or how poorly the simple pocket formula approximates double parton scattering. Indeed, several measurements of $\sigma_{eff}$ has been made, and vary between $5-20$~mb, as shown in Figure~\ref{fig:sigma_eff}. With a DPS interpretation of the LHCb measurement  \cite{Aaij:2012dz,Seymour:2013sya}, the results vary in an even broader range. Great care must however be taken, because in order to separate double from single parton scattering the characteristic kinematical distribution of the final state particles is used. Several of the approximations made to arrive at the pocket formula can have sizable effects on the distributions in double partons scattering which leads to a risk of misjudging the DPS contribution.

Despite its shortcomings, the pocket formula approximation to DPS is used in many phenomenological studies. The reason for this, is the difficulties that arise when moving away from the simple picture, and treat DPS with a more solid base in perturbative QCD, see for example \cite{Diehl:2011tt,Manohar:2012jr,Blok:2013bpa,Snigirev:2014eua}. Apart from the approximation mentioned above, a thorough treatment of DPS requires dealing with different types of correlations between two partons inside a proton. On one hand, this opens up the interesting possibility of extracting information about the connections between the quarks and gluons inside a proton and a rich phenomenology. On the other hand, it leads to several complications which have to be dealt with. 

\section{Polarization in double parton scattering}
There are several types of correlations between two partons inside a proton which break the simplified ansatz and which are not made explicit in \eqref{eq:x-sec}. These include different types of correlations between the quantum numbers of the two partons, see e.g. \cite{Diehl:2011tt,Kasemets:2014yna} for a more detailed account of the different possibilities.
One interesting type of correlation is between the spin of the two partons. This leads to polarized double parton distributions (DPDs) in the cross section formula \eqref{eq:x-sec}. Spin correlations has the interesting property that they are directly and in a calculable way linked with the distributions of the final state particles, and it is therefore possible to study how such correlations affects measurements at the LHC. 

There has so far only been a handful of studies in DPS which includes the possibility of spin correlations. The different possible DPDs were written down in \cite{Diehl:2011yj,Manohar:2012jr}. How they appear in cross section calculations was studied in detail in double Drell-Yan process  \cite{Manohar:2012jr,Kasemets:2012pr}, including production of $W$ bosons, and in double $c\bar{c}$ production in \cite{Echevarria:2015ufa}. Spin correlations was found to be sizable quark models \cite{Chang:2012nw,Rinaldi:2014ddl}. The polarized distributions can be divided into two types, longitudinal and transverse/linear polarization, where transverse is for quarks and linear is for gluons. The longitudinal polarization affects the total size of double parton scattering cross sections as well as the distributions of final state particles in, for example, rapidity and $p_T$. The transverse/linearly polarizations produce azimuthal modulations, leading to azimuthal asymmetries. For example in double Drell-Yan, the transversely polarized quarks result in a $\cos(2\phi)$ term, where $\phi$ is the azimuthal angle between the outgoing leptons (as opposed to anti-leptons) of the two hard interactions. As such, it manifests a direct correlation between the outgoing directions of the particles from one of the interactions to the outgoing directions of the particles produced in the second.

Quantifying such statements requires an ansatz for the unpolarized as well as polarized DPDs in the cross section expression. Since these distributions has not yet been measured, one must resort to models. With the precise knowledge of the PDFs it is natural to start building such a model for the unpolarized DPDs from the product of two PDFs, possibly modified to simulate different types of correlations and to satisfy momentum conservation. The polarized distributions (describing parton-parton spin correlations) does not have any single parton analogue and their modeling must follow a different route. The polarized distributions should satisfy positivity bounds, derived from positivity or probability interpretation \cite{Diehl:2013mla}. In order to obtain an indication of how large an effect the spin correlations can have on the level of the cross section, one can saturate these bounds at a low scale, and then evolve them via double DGLAP evolution to the scale of the process. The evolution washes out the spin correlations between the two partons, but the rate, and thus the final effect, has strong dependence both on the type of the partons, their $x_i$ fractions and the type of polarization considered. 

\begin{figure}[t]
\begin{center}
\includegraphics[width=0.35\textwidth]{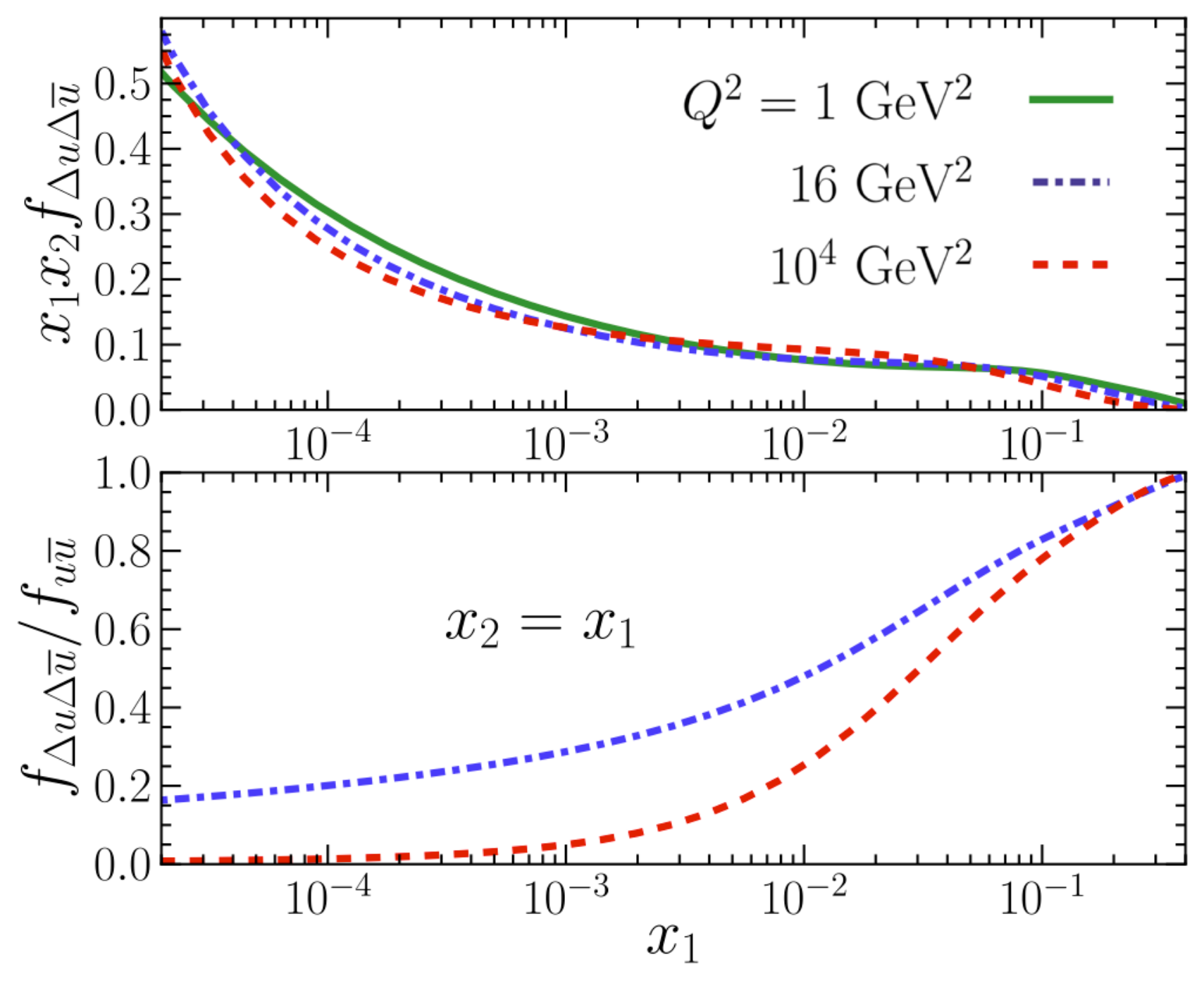}
\quad
\includegraphics[width=0.35\textwidth]{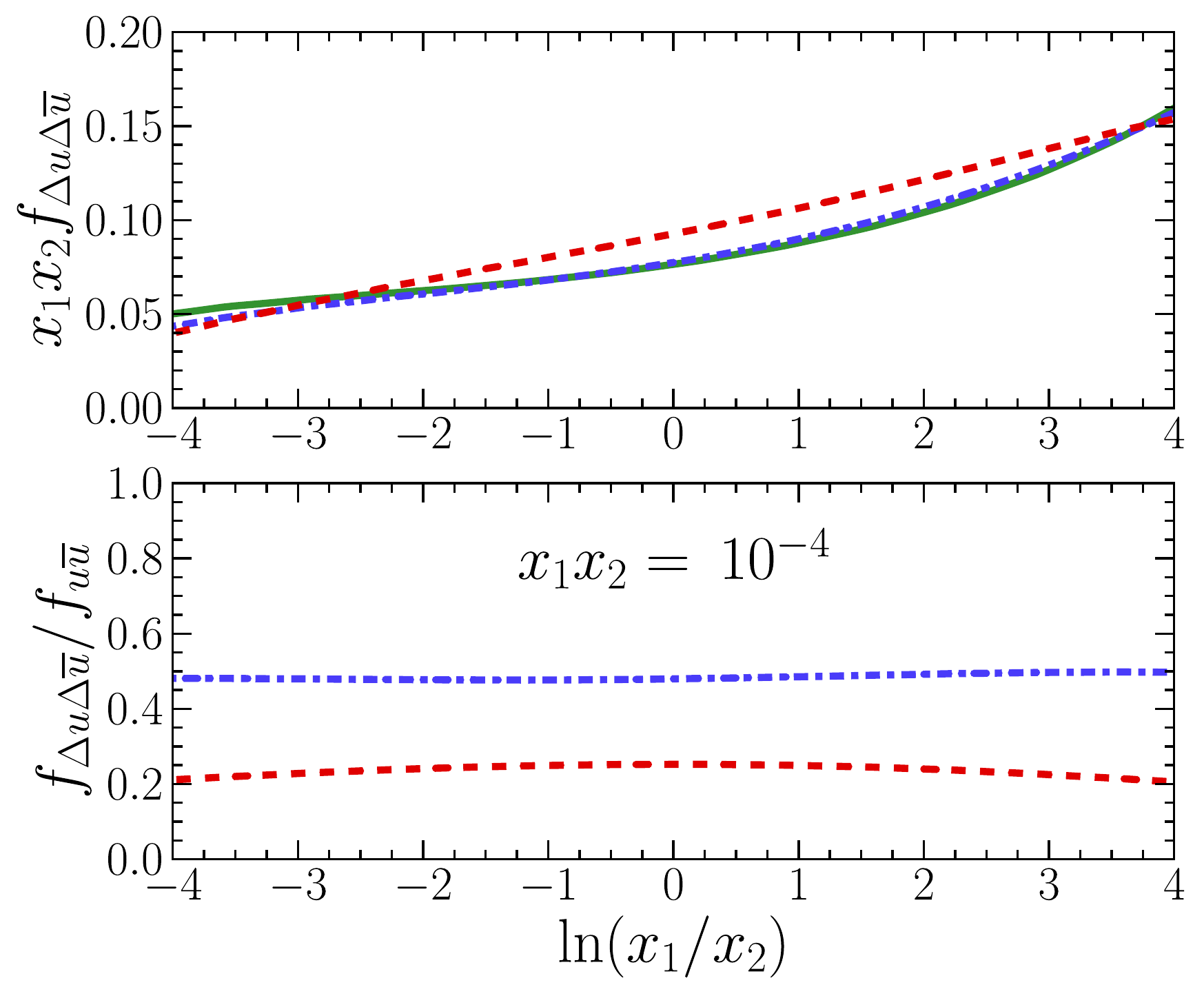}
\end{center}
\caption{\it
  Effect of double DGLAP evolution on the longitudinally polarized DPD for an up and an anti-up quark, starting from maximal polarization at the initial scale ($Q_0=1$~GeV). The lower panels show the degree of polarization.
}
\label{fig:pol_evo1}
\end{figure}

\begin{figure}[t]
\begin{center}
\includegraphics[width=0.35\textwidth]{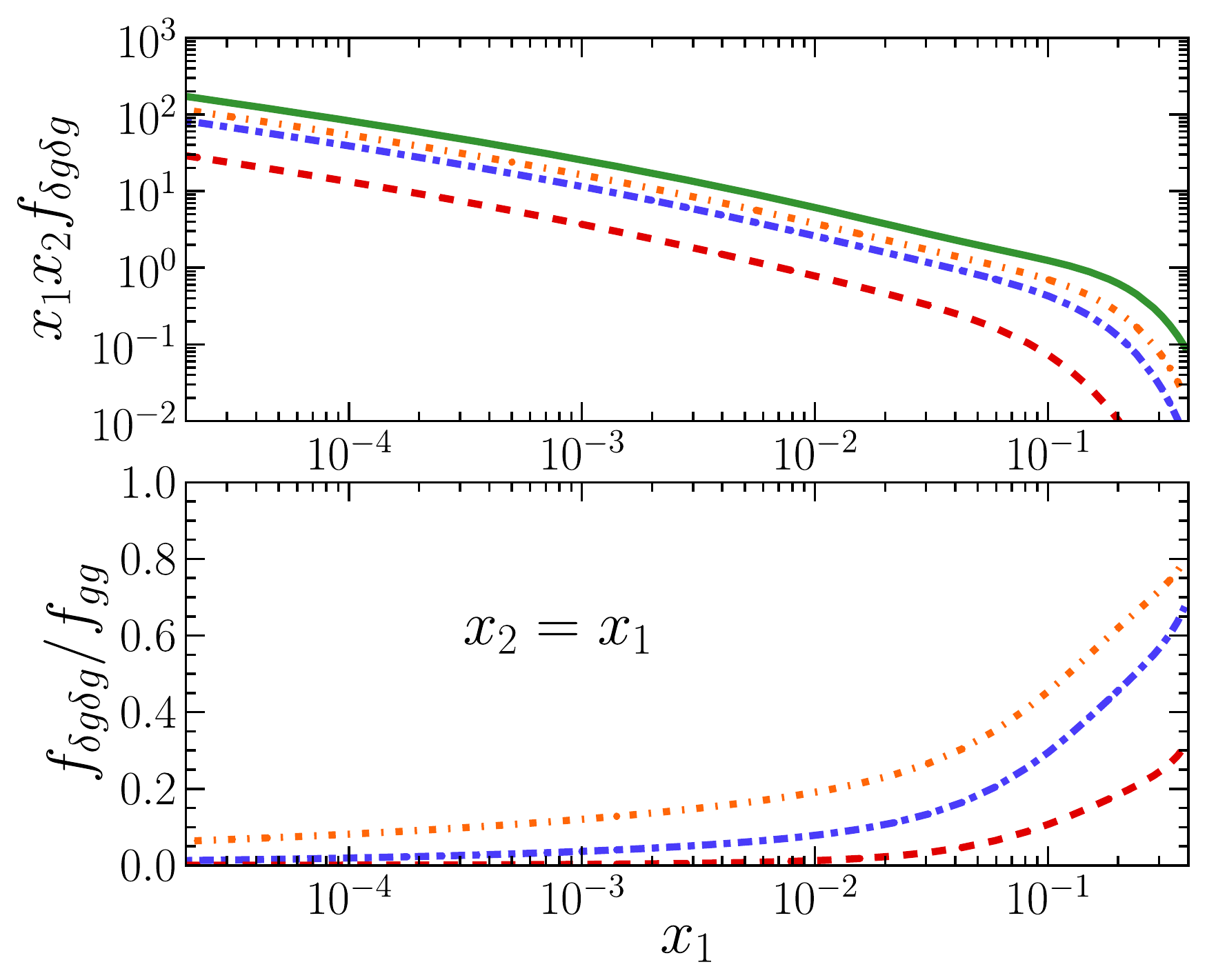}
\quad
\includegraphics[width=0.35\textwidth]{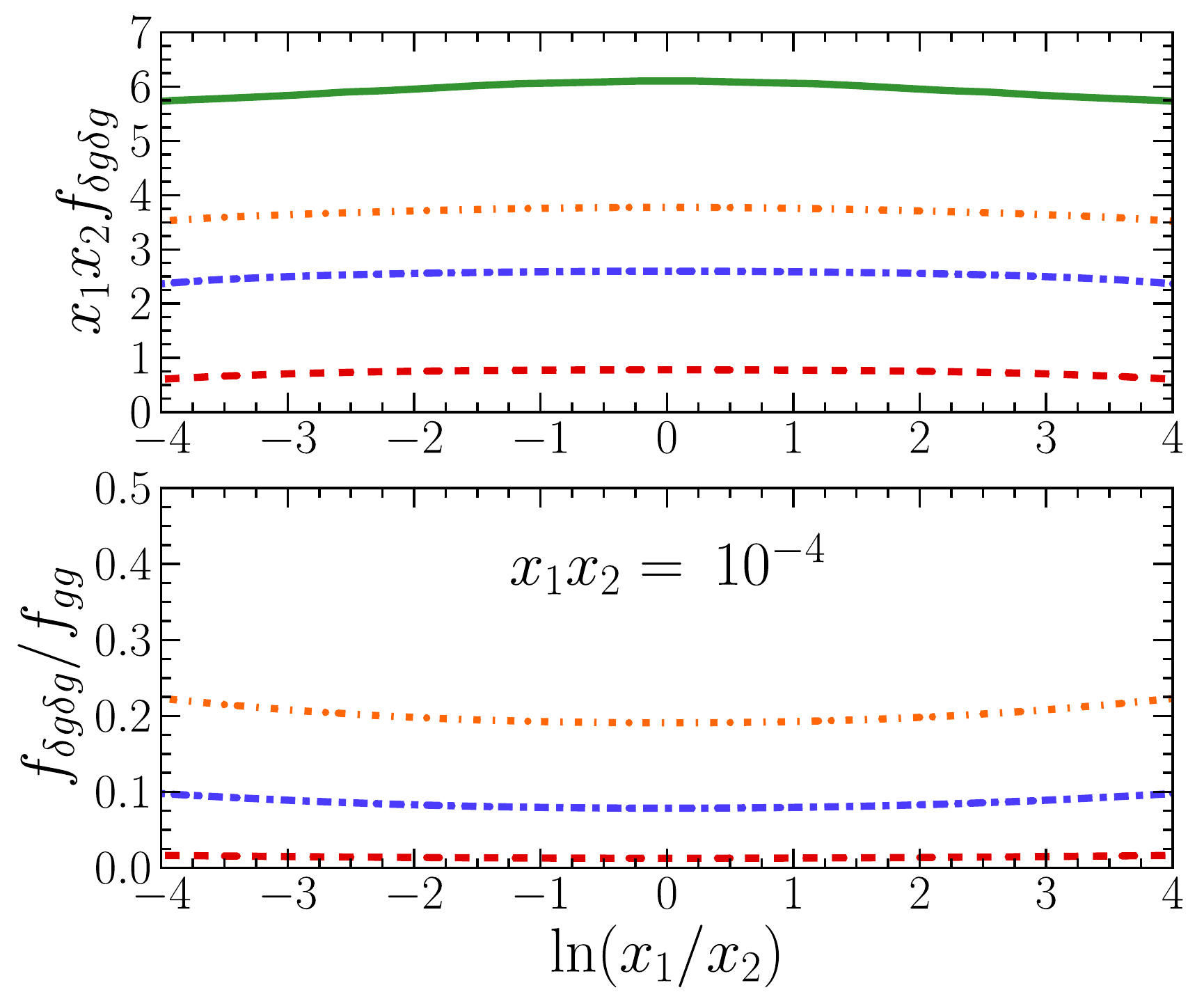}
\end{center}
\caption{\it
  Effect of double DGLAP evolution on the linearly polarized DPD for two gluons, starting from maximal polarization at the initial scale ($Q_0=1$~GeV). The lower panels show the degree of polarization.
}
\label{fig:pol_evo2}
\end{figure}

Generically quark polarizations tend to be washed out much slower than gluon polarization, and longitudinal polarizations remains large up to higher scales than transverse/linear polarizations. As an example, we demonstrate the effect of double DGLAP evolution on the DPD for a longitudinally polarized up and anti-up in Figure~\ref{fig:pol_evo1}. The staring conditions are chosen such that the positivity bounds are saturated at an initial scale of $Q_0=1$~GeV, leading to 100\% polarization. The upper panels show the change of the distributions themselves, while the lower show the impact of evolution on the  ratio of the polarized to the unpolarized distribution, i.e. on the degree of polarization. We can see, that even at large scales the polarized contribution can remain sizable and, for example, still be above $20\%$ at $Q^2=m_Z^2$. For gluons, in particular at low $x_i$ and for the linearly polarized distribution, the polarization rapidly approach zero with increasing evolution scale. Figure \ref{fig:pol_evo2} shows the double linearly polarized gluons, the type of polarization which is most rapidly washed out. Already at low scales, the polarization is in essence absent for a large range in $x_i$. The main source of this suppression is the very rapid increase in the unpolarized double gluon distribution driven by the $1/x$ term in the gluon DGLAP splitting kernel.

\begin{figure}[t]
\begin{center}
\includegraphics[width=0.32\textwidth]{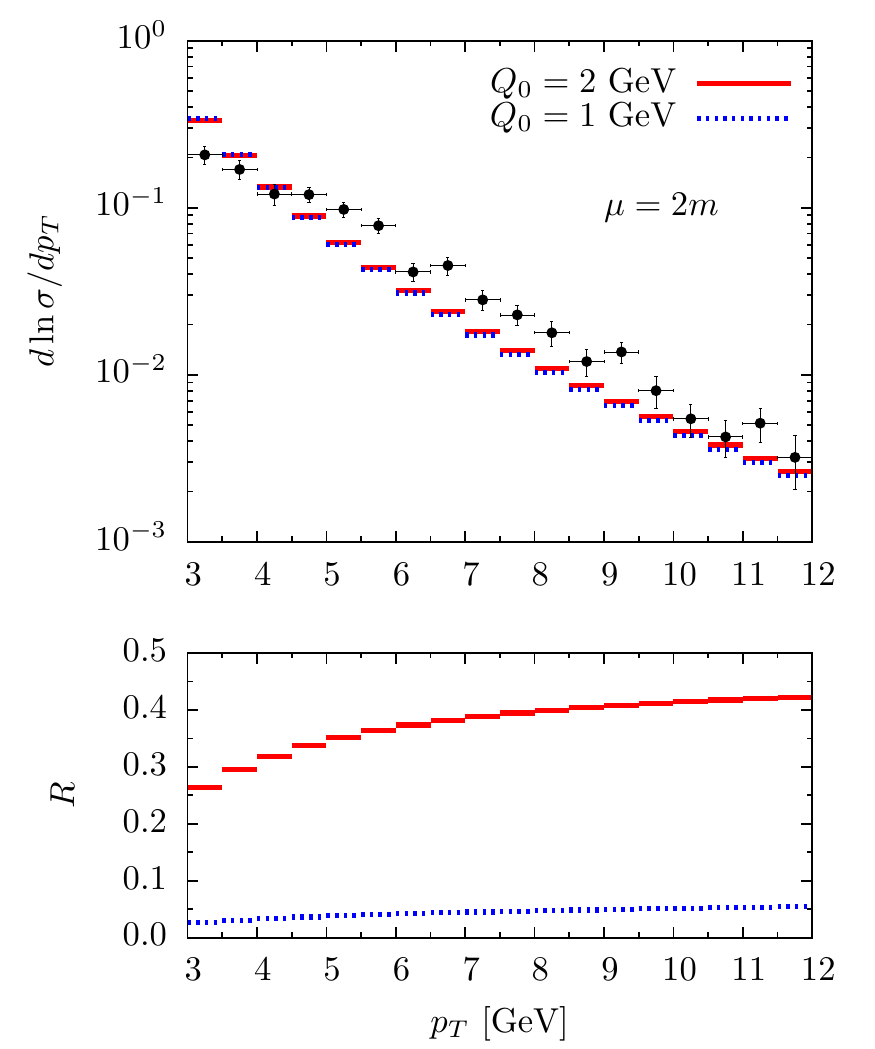}
\includegraphics[width=0.32\textwidth]{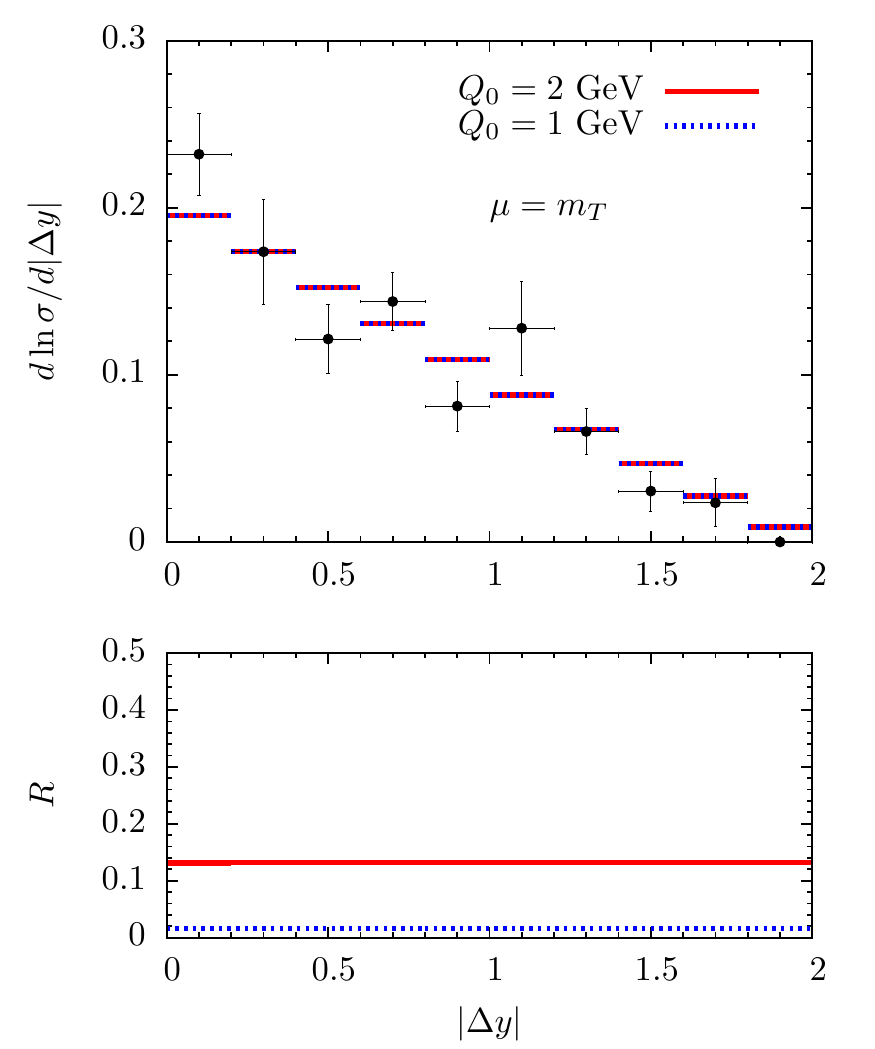}
\includegraphics[width=0.32\textwidth]{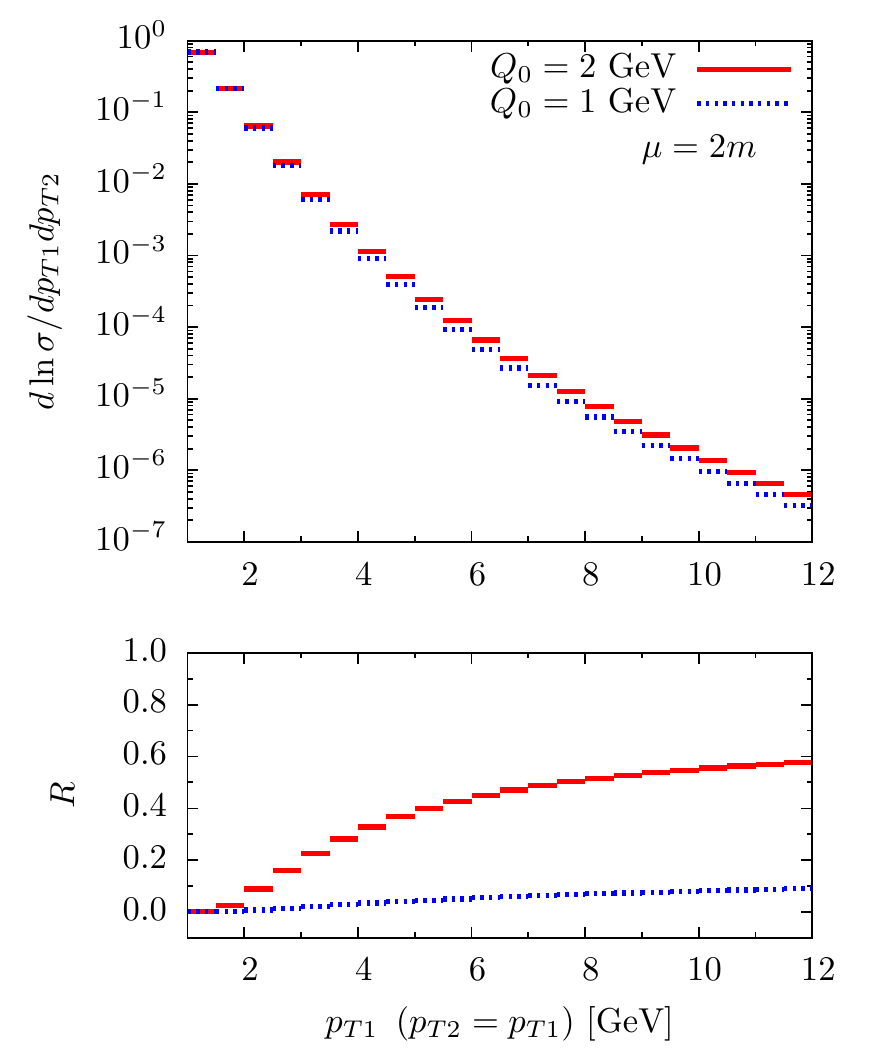}
\end{center}
\caption{\it
  Normalized cross section vs the transverse momentum of one of the charm quarks (left) and rapidity difference (middle)  -- overlaid with the LHCb $D^0D^0$ data \cite{Aaij:2012dz}, and vs the transverse momentum of both charm quarks (right). The lower panels show the relative size of the polarized contribution. 
}
\label{fig:open_charm_LHCb}
\end{figure}

So far, there has only been one quantitative study of the polarization effects at the cross section level. This was in the production of double open charm, or double $D_0D_0$ mesons, in LHCb kinematics \cite{Echevarria:2015ufa}. This process is of particular interest for DPS, since it has been shown that DPS dominates over SPS \cite{vanHameren:2014ava,vanHameren:2015wva}. It was demonstrated that polarization can have a large impact on the size of the cross section, but did not induce any large shape dependence for the distributions measured so far by the LHCb collaboration. This can be seen from the two first panels of Figure~\ref{fig:open_charm_LHCb}, where the calculation is overlaid with the LHCb data. Care must be taken in comparing the results, since the calculation is for double open charm, and the data for double $D_0$. For the normalized cross section in Figure~\ref{fig:open_charm_LHCb}, this should not affect the results at the current level of precision. We can see that there is good agreement between the data and the DPS prediction. In the lower panels, we demonstrate the relative contribution of the polarized over the unpolarized production for two different staring scales $Q_0=1,2$~GeV and for two choices of scale for the process $\mu = 2m_c$ and $\mu=m_T$. We can see that the polarized contribution has a strong dependence on the initial scale, since lower initial scale means more evolution, and thus less  remaining polarization. The shape of the polarized and unpolarized result is rather similar, and it is hence difficult two distinguish the contributions. If we however turn to the double differential distribution in the rightmost panel, we can see a strong relative increase in the polarized contribution towards larger $q_T$, for $\mu=2m_c$ and $Q_0=2$~GeV. Such shape differences in the cross section could, once confronted with measurements, lead to first experimental indications, or limits, on polarization in double parton scattering.

Many aspects in the description of polarization in DPS closely resembles the description of polarization in transverse momentum dependent single parton scattering. The presence a transverse vector ($k_T$ in TMDs and $\y$ in DPDs) and the possibility to have two-particle spin correlations (proton-parton vs parton-parton) leads to similar decompositions of the hadronic matrix elements \cite{Boer:1997nt,Mulders:2000sh,Diehl:2011yj}. This is reflected in cross section calculations, as in both cases, distributions for transversely polarized quarks and linearly polarized gluons give rise to azimuthal asymmetries. While for TMDs, much effort both on theoretical and experimental sides, has been devoted to the study of the asymmetries, the landscape of spin-asymmetries in DPS remains largely unexplored.

\subsection*{Acknowledgements}
We acknowledge financial support from the European Community under the "Ideas" program QWORK (contract 320389).

\end{document}